\documentclass[a4paper,fleqn,usenatbib]{mnras}

\usepackage{newtxtext,newtxmath}

\usepackage[T1]{fontenc}
\usepackage{ae,aecompl}

\usepackage{graphicx}   
\usepackage{amsmath}    
\usepackage{amssymb}    

\def\eg{{\it e.g.,}\thinspace}
\def\eq{\begin{equation}}
\def\en{\end{equation}}

\def\lesssim{\raisebox{-0.3ex}{\mbox{$\stackrel{<}{_\sim} \,$}}}
\def\gtrsim{\raisebox{-0.3ex}{\mbox{$\stackrel{>}{_\sim} \,$}}}
\def\deg{\hbox{$^{\circ}$}}

\def\apj{{\it Ap.J.}}

\def\apjs{{\it Ap.J. Suppl.}}
\def\aap{{\it A\&A}}
\def\aaps{{\it A\&A Suppl.}}
\def\mnras{{\it MNRAS}}
\def\P3hat{{\mathaccent 94 P}_3}

\def\nat{{\it Nature}}

\title[Periodic Longitude-Stationary Non-Drift Emission]{Periodic Longitude-Stationary Non-Drift Emission in Core-Single Radio Pulsar B1946+35}
\author[Dipanjan Mitra \& Joanna Rankin]{Dipanjan Mitra$^{1,2,3}$ \& Joanna Rankin$^{1}$ \\
$^1$Physics Department, University of Vermont, Burlington, VT 05405\thanks{dmitra@uvm.edu; Joanna.Rankin@uvm.edu} \\
$^2$National Centre for Radio Astrophysics, Ganeshkhind, Pune 411 007 India\thanks{dmitra@ncra.tifr.res.in} \\
$^3$Janusz Gil Institute of Astronomy, University of Zielona G\'ora, ul. Szafrana 2, 65-516 Zielona G\'ora, Poland}

\date{In original form year month day}
\pubyear{year}
\begin{document}
\label{firstpage}
\pagerange{\pageref{firstpage}--\pageref{lastpage}}
\maketitle

\begin{abstract}
Radio pulsar PSR B1946+35 is a classical example of a core/cone triple pulsar
where the observer's line-of-sight cuts the emission beam centrally.  In this paper 
we perform a detailed single-pulse polarimetric analysis of B1946+35 using
sensitive Arecibo archival and new observations at 1.4 and 4.6 GHz to re-establish 
the pulsar's classification wherein a pair of  inner conal ``outriders'' surround a 
central core component.  The new 1.4 GHz observation consisted of a long 
single pulse sequence of 6678 pulses, and its fluctuation spectral analysis revealed that the pulsar 
shows a time-varying amplitude modulation, where for a thousand periods or so 
the spectra have a broad low frequency ``red'' excess and then at intervals they 
suddenly exhibit highly periodic longitude-stationary modulation of both the core 
and conal components for several hundred periods. The fluctuations of the 
leading conal and the core components are in phase, while those in the trailing 
conal component in counterphase. These fluctuation properties are consistant
with shorter pulse sequence analyses reported in an earlier study by 
\citet{2006A&A...445..243W,2007A&A...469..607W}
as well as in our shorter pulse sequence data sets. 
We argue that this dual modulation of core 
and conal emission cannot be understood by a model 
where subpulse modulation is associated with the plasma {\bf E}$\times${\bf B} drift phenomenon.  
Rather the effect appears to represent a kind of periodic emission-pattern change 
over timescales of $\sim$18 s (or 25 pulsar periods), which has not been reported previously 
for any other pulsar. 
\end{abstract}


\maketitle


\section{Introduction}
\label{sec1}

Coherent radio emission from pulsars is thought to be generated by the nonlinear 
growth of relativistic plasma instabilities \citep[e.g.][]{1995JApA...16..137M,2004ApJ...600..872G}.
This emission is known to originate from regions of open dipolar 
magnetic field in the inner magnetosphere \citep[e.g.][]{1991ApJ...370..643B,2004A&A...421..215M}. 
Such emitting regions of open, outwardly curving field are 
roughly circular or conical in shape and produce a beam of radius $\rho$ (see {\it 
Empirical Theory of Pulsar Emission} series: \citealt{1993ApJS...85..145R,1993ApJ...405..285R}, 
hereafter ET VIa,b; 
\citealt{1999A&A...346..906M}).  The average emission profiles of individual radio 
pulsars are highly stable in shape and reflect the manner in which our sightline 
samples a rotating pulsar's beam.  A large variety of different profile forms are 
observed among the pulsar population, consisting of one to a usual maximum of 
five components.  A model wherein pulsar emission beams are comprised by a 
central core beam and/or two nested conal beams---and sampled by sightlines 
cutting centrally or obliquely for different pulsars---can largely accommodate the 
observed profile forms (see ET I, IV and VIa). 

Pulsar magnetospheres can also be studied using a pulsar's individual pulses which  
show significant variability on various time scales.  At extremely short timescales, 
structures on nanosecond or microsecond scales \citep[e.g.][]{2003Natur.422..141H,
1979AuJPh..32....9C,2015ApJ...806..236M}
are thought to be related to the nonstationary 
behaviour of the emitting plasma.  The phenomenon of subpulse drifting is revealed 
on timescales of several tens of seconds and is possibly related to plasma dynamics 
seen as {\bf E}$\times${\bf B} drift in the pulsar magnetosphere (\citealt{1970Natur.227..692B};
\citealt{1975ApJ...196...51R}, hereafter RS75).  The least understood 
phenomena are mode-changing and nulling of pulsar signals, which are associated 
with state changes in the pulsar emission and occur on time scales of minutes to hours 
\citep[e.g.][]{1986ApJ...301..901R,2007MNRAS.377.1383W}
to even months or years as in 
the case of rotating radio transients or RRATs \citep{2006Natur.439..817M} and 
intermittent pulsars \citep[e.g.][]{2006Sci...312..549K}.  Currently there are no theoretical 
models that can explain the phenomenon of pulsar emission overall, and hence 
clues from new observations and phenomenology are essential.

There are, however, strong relationships between average-profile and drifting-subpulse 
properties that have been identified, which are of fundamental interest to our work
below.  In order to fully understand these relationships, let us briefly review the logic of 
profile classification that permits us to infer a pulsar's beam and basic quantitative 
emission geometry in terms of the magnetic colatitude $\alpha$ and sightline impact 
angle $\beta$.  These angles can usually be estimated by interpreting the linear polarization 
position-angle (PPA) traverse across the profile in terms of the rotating-vector model 
(RVM, \citealt{1969ApL.....3..225R,1970Natur.225..612K}).  This function exhibits an 
`S' shape as the sightline encounters a range of dipolar magnetic field planes and is a 
strong function of $\alpha$ and $\beta$, such that its slope close to the inflection point 
$R_{\rm pa} = \sin\alpha/\sin\beta$.  Shallower traverses or smaller $R_{\rm pa}$ values indicate 
that the sightline cuts the emission beam tangentially and samples the conal emission 
whereas steeper traverses or larger $R_{\rm pa}$s show that the sightline cuts centrally and 
hence samples both the core and conal parts of the overall pulsar beam.  As per the 
classification scheme in ET I, IV and VI, two different types of pulsars with single profiles 
are observed in the frequency band around or below 1 GHz.  These can be distinguished 
according to how their profile forms evolve at higher or lower frequencies.  One type 
broadens into a double structure at lower frequencies; it is designated ``single of the 
double type'' {\bf S$_d$} because of its close relationship to pulsars with conal double 
{\bf D} profiles.  Whereas, the second type, of most interest here, remains single at low 
frequencies but develops a pair of conal ``outrider'' components at higher frequencies, 
so is designated ``single of the triple type'' {\bf S$_t$} because of its close relation to 
pulsars with core/cone triple {\bf T} profiles (\citealt{1983ApJ...274..333R}, hereafter ET I). 

Conal single {\bf S$_d$} pulsars exhibit the remarkable drifting-subpulse phenomenon, but 
the effect has never been seen in core-single {\bf S$_t$} pulsars.  Drifting subpulses are manifested 
within a pulsar's sequences when orderly longitude motion occurs from pulse to pulse, 
producing ``drift bands'' with a well defined drift frequency $f_3$ (or period $P_3 = 1/f_3$) 
and 
longitude separation $P_2$.  Longitude-resolved fluctuation spectra are employed to 
recover the $f_3$ amplitude and phase by performing fast Fourier transforms along each 
longitude of the pulse sequence.  For {\bf S$_d$} pulsars $f_3$ can often be determined with 
precision, with its phase having either a positive or negative slope across the profile.  Well 
known examples of {\bf S$_d$} pulsars are PSR's B0031--07, B0943+10, B0809+74, B1944+17, 
etc., and other examples can be found in Table 2 of ET III (Rankin 1986), and in the major 
subpulse drift studies by \citet[][WES06 and WES07 hereafter]{2006A&A...445..243W,2007A&A...469..607W} and 
\citet{2016ApJ...833...29B}.  {\bf S$_d$} pulsars typically represent an older population of radio pulsars with 
smaller slowdown energy rates $\dot{E} \lesssim 10^{32}$ ${\rm ergs\,s^{-1}}$ 
\citep[ET III][]{2000ApJ...541..351G}.

\begin{figure*}
\begin{tabular}{cc}
\mbox{\includegraphics[height=80mm,width=70mm,angle=-90.]{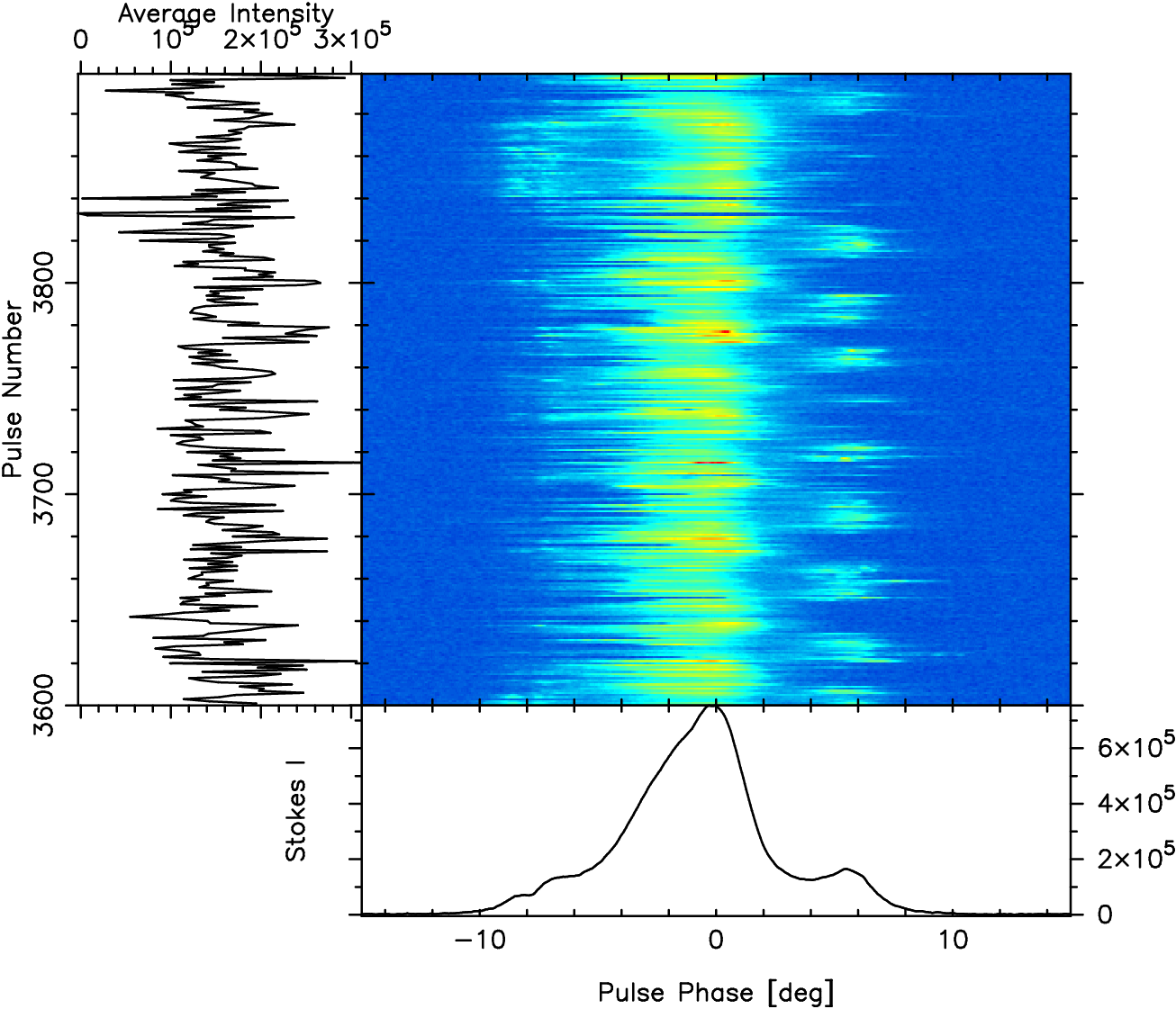}} &
\mbox{\includegraphics[height=80mm,width=70mm,angle=-90.]{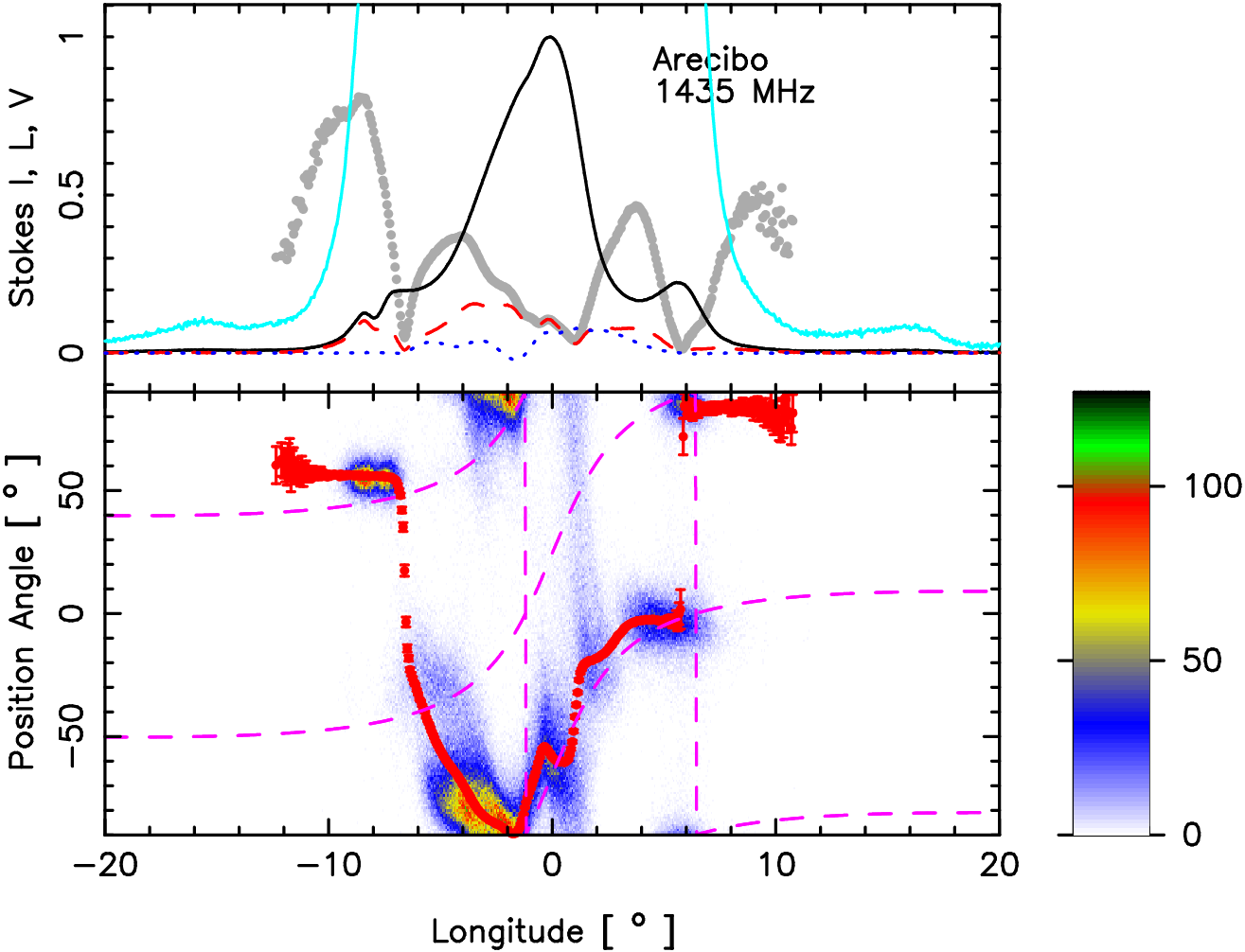}}\\
\mbox{\includegraphics[height=80mm,width=70mm,angle=-90.]{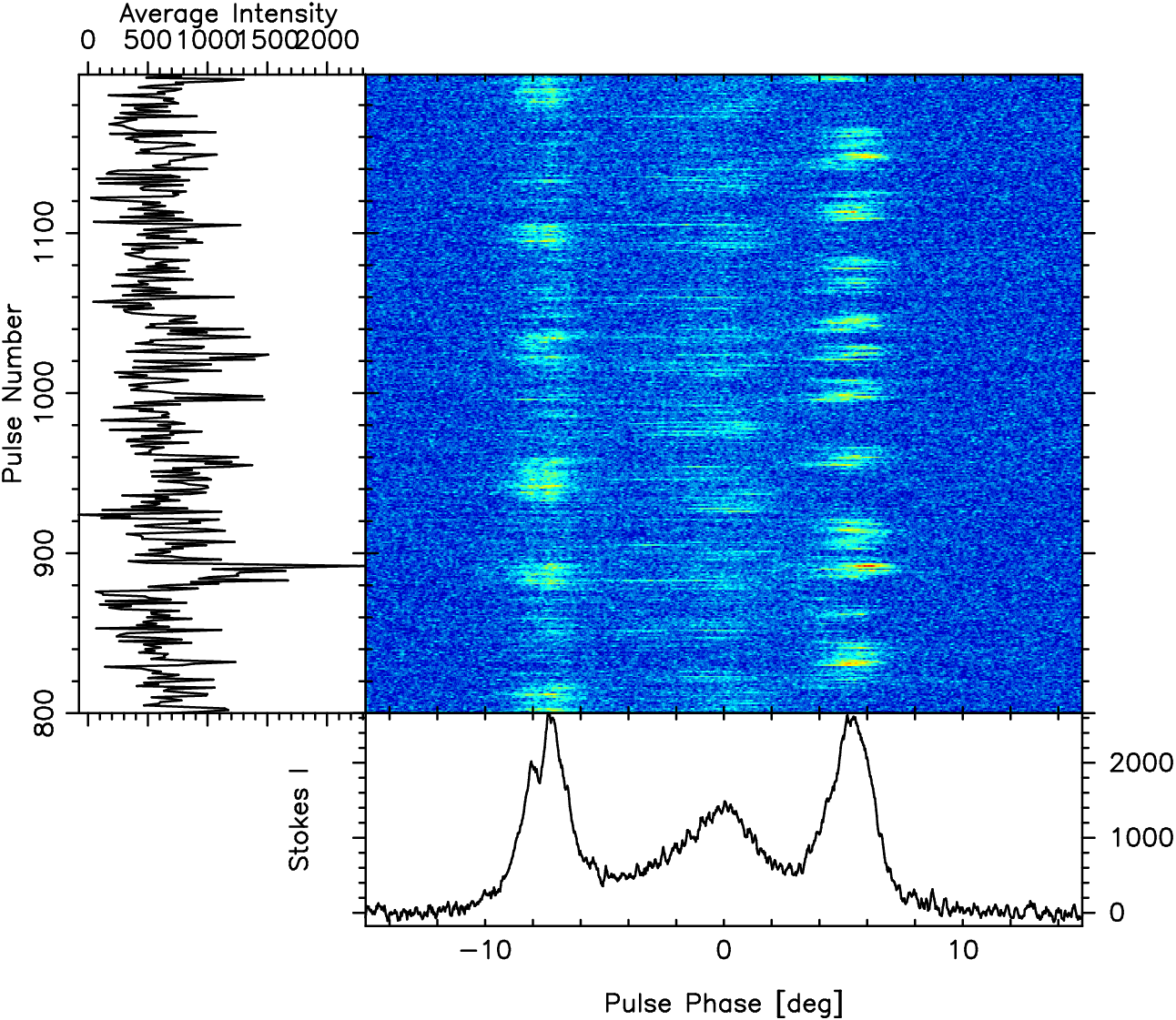}} &
\mbox{\includegraphics[height=80mm,width=70mm,angle=-90.]{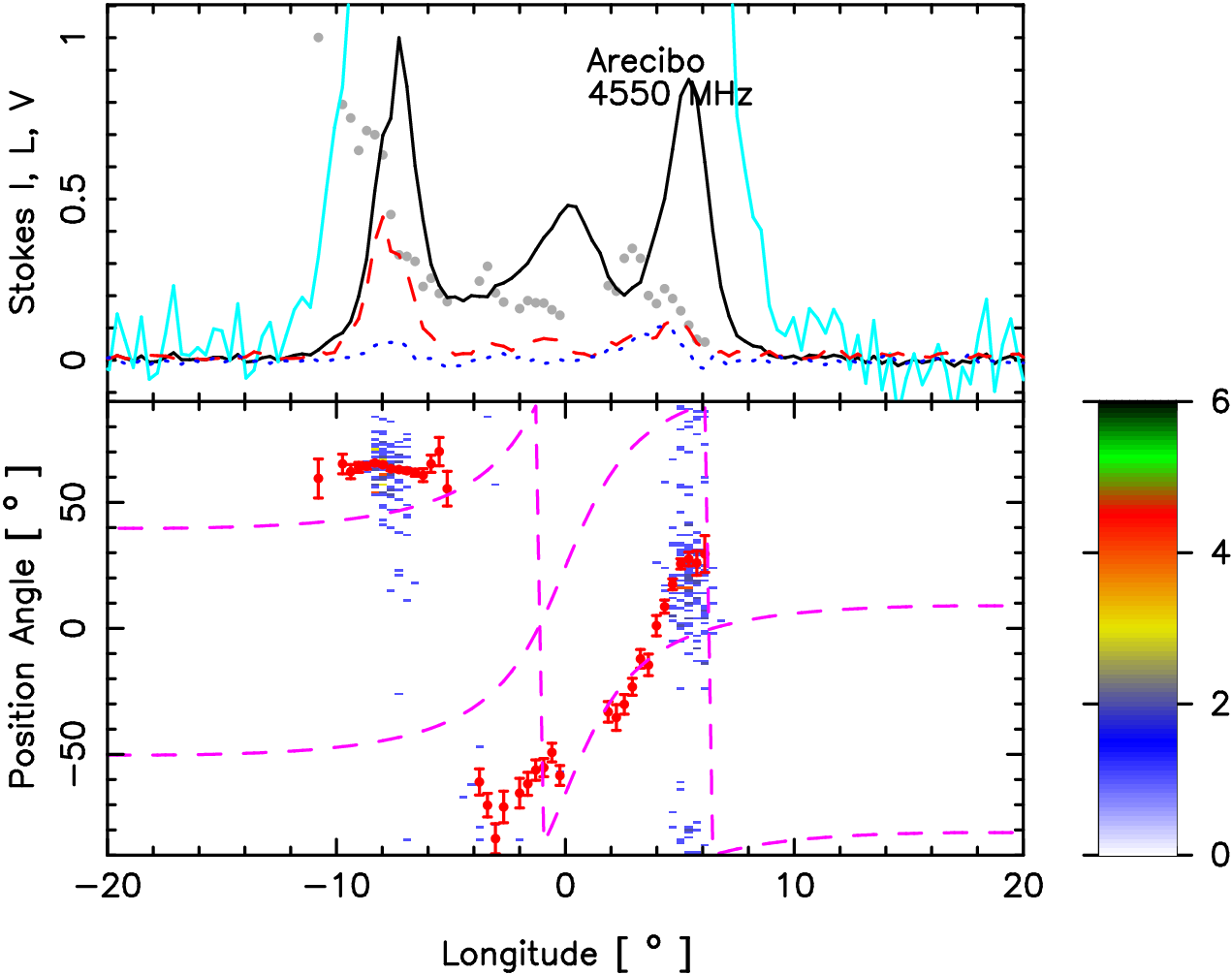}}\\
\end{tabular}
\caption{PSR B1946+35 pulse sequences (left hand displays) and average polarization 
characteristics (right hand displays) at 1.6 (LSP3) and 4.6 GHz (CPS1), respectively.  
The latter's upper panels give the total intensity (Stokes $I$; solid curve), the total linear 
($L$ [=$\sqrt{Q^2+U^2}$]; dashed red), the circular polarization (Stokes $V$; dotted blue), 
the fractional linear polarization ($L/I$; gray points) and a zoomed total intensity (10$\times I $; 
cyan curve).  The PPA [=${1\over 2}\tan^{-1} (U/Q)$] densities (lower panels) within each 
1$\degr$x1$\degr$-sample cell, corresponding to samples larger than 2 $\sigma$ in $L$, are 
plotted according to the color bars at the lower right, and have been derotated to infinite 
frequency.  The average PPA traverses (red) are overplotted, and the RVM fit is plotted 
twice for the two polarization modes (magenta dashed) on each panel.  The plot origins 
are taken at the fitted PPA inflection point.}
\label{fig1}
\end{figure*}

\begin{table*}
\begin{center}
\caption{Details of the B1946+35 pulse sequences observed with the Arecibo telescope.  
The first four columns give the observing frequency, MJD and the 
Arecibo Project-ID, the raw time resolution, and the 
bandwidth. The fifth column gives the pulse-sequence range over which overlapping FFTs 
(of length OFFT in column 8) have been performed.  The outside half-power pulse width 
measured for the full pulse sequence and the beam radius $\rho$ are given in columns 6 
and 7.  The fluctuation-frequency values $f_3^{offt}$ and $Q$ factor obtained for the 
corresponding range of pulses are given in columns 9 and 10.}
 \begin{tabular}{ccccccccccc}
 \hline
 \hline
Band  & MJD/AO ProjID   & Tres & BW & Range    & Width & $\rho$& OFFT & $f_3^{offt}$ & $Q = \frac{f_3^{offt}}{\Delta f}$\\
(GHz) &       & ($\mu$s)     & MHz    & (pulses) & (\deg) & (\deg) &  & (c$P^{-1}$) &                          \\
\hline
1.4   & 52837/P1734 & 1024 & 400&    1--1366  & 15.6$\pm$0.1  & 4.6$\pm$0.1 &256  & 0.0202$\pm$0.015  & 0.83             \\
      & (LPS1)&      &    &  512--1262  & & &256  & 0.0194$\pm$0.002  & 3.40             \\
 &       &      &    &  &             & & &                          \\
1.4   & 55632/P2532 & 512  & 400&    1--1031  & 15.4$\pm$0.1  & 4.6$\pm$0.1 &256  & 0.0238$\pm$0.002  & 1.45              \\
      & (LPS2)&      &    &  512--1024  & & &256  & 0.0239$\pm$0.002  & 2.93              \\
 &       &      &    &  &             & & &                          \\
1.4  & 57638/P3031 & 120  & 400&  1--6678 & 15.5$\pm$0.1   & 4.6$\pm$0.1&   256  & 0.0208$\pm$0.014  & 0.60              \\
      & (LPS3)&      &    & 1250--1761 & & & 256  & 0.0193$\pm$0.003  & 3.21             \\
      &       &      &    & 2500--3011 & & & 256  & 0.0235$\pm$0.003  & 3.85           \\
      &       &      &    & 3550--4062 & & & 256  & 0.0312$\pm$0.002  & 6.11           \\
      &       &      &    & 5100--5612 & & & 256  & 0.0314$\pm$0.002  & 6.63           \\
 &       &      &    &  &             & & &                          \\
4.6   & 57284/P2995 & 51   & 1187&   1--1237 & 15.1$\pm$0.1  &4.5$\pm$0.1 &256  & 0.0104$\pm$0.029  & 0.34           \\
      & (CPS1)&      &    &  520--776   & & &256  & 0.0184$\pm$0.002  & 2.04              \\
\hline
\hline
\end{tabular}
\end{center}
Note: Basic parameters for PSR B1946+35 are : dispersion measure DM = 129.07 ${\rm pc\,cm^{-3}}$, rotation measure RM=116 ${\rm rad\,m^{-2}}$, \\period $P$ = 0.717 s, and period derivative $\dot{P} = 7.061 \times 10^{-15}$${\rm s\,s^{-1}}$.
\label{tab1}
\end{table*}

The fluctuation spectra of the core features in {\bf S$_t$} pulsars, on the other hand, show very 
different pulse-sequence modulation properties.  ET III had established that the fluctuation 
spectra of most core-single pulsars were largely featureless.  More recent analyses with 
more sensitive observations and analysis techniques largely reiterate this conclusion, 
though there is some evidence that in a few cases {\bf S$_t$} pulsars exhibit diffuse fluctuation 
features (\eg B0136+57, B0823+26, B1642--03 or B2255+58; see WES06, WES07; \citet{2016ApJ...833...29B}) 
with no discernible longitude motion, consistent with phase-stationary amplitude 
modulation.  The conal components of pulsars with cores sometimes have sharper and 
steadier fluctuations with well defined $f_3$; however the modulation tends to be stationary 
in phase across the conal components.  Importantly, pulsars with {\bf S$_t$} profiles represent 
a younger population with slowdown rates $ \dot{E} \gtrsim 10^{32}$ ${\rm ergs\,s^{-1}}$.

Here, we report an unusual type of pulse-sequence (hereafter PS) modulation in pulsar 
B1946+35.  The pulsar is moderately bright and core-dominated when observed 
at L-band (1.4 GHz) but develops conal outriders at higher frequencies \citep[e.g.][]{1988MNRAS.234..477L}. 
PSR B1946+35 has been classified as having an {\bf S$_t$} profile 
in ET IV and VI.  Interestingly, single pulse fluctuation spectra studies at P-band (327 MHz) and 
L-band by WES06 and WES07 indicate strong modulation features across its pulse profiles.  
Furthermore, the studies also find very different $P_3$ values of 55$\pm$9$P$ (or $f_3 \sim 0.018$ cycles $P^{-1}$ 
(hereafter c$P^{-1}$) where $P$ is the pulsar period) 
and 33$\pm$2 $P$(or $f_3 \sim 0.030$ c$P^{-1}$) at P-band and L-band respectively, which the authors suggest
may be due to the relatively short lengths of the observations that were perhaps insufficient 
to accurately estimate the average $P_3$ value.

In this paper, we use both archival and recent Arecibo observations to carry out a detailed 
study of this pulsar on a single pulse polarimetric basis.  As we will illustrate, the pulsar's 
fluctuation spectra suggest a highly periodic coupled modulation of both the core and conal 
emission, which is unusual for an  {\bf S$_t$} pulsar.  We conclude that the pulsar's modulation 
cannot be understood as subpulse drift within the usual rotating carousel subbeam model of 
RS75.  How we reach this conclusion is the argument of this paper.  In \S\ref{obs} we discuss 
the observations. \S\ref{sec3} assembles the evidence pertaining to B1946+35's profile 
classification and reviews the logic of our understanding it to be a core-single {\bf S$_t$} pulsar.  
\S\ref{sec4} discusses the fluctuation spectra and compares the results with known subpulse 
drift phenomena, and \S\ref{sec5} gives the conclusion of this work. 
    
\begin{figure*}
\begin{tabular}{cc}
\mbox{\includegraphics[height=70mm,width=60mm,angle=-90.]{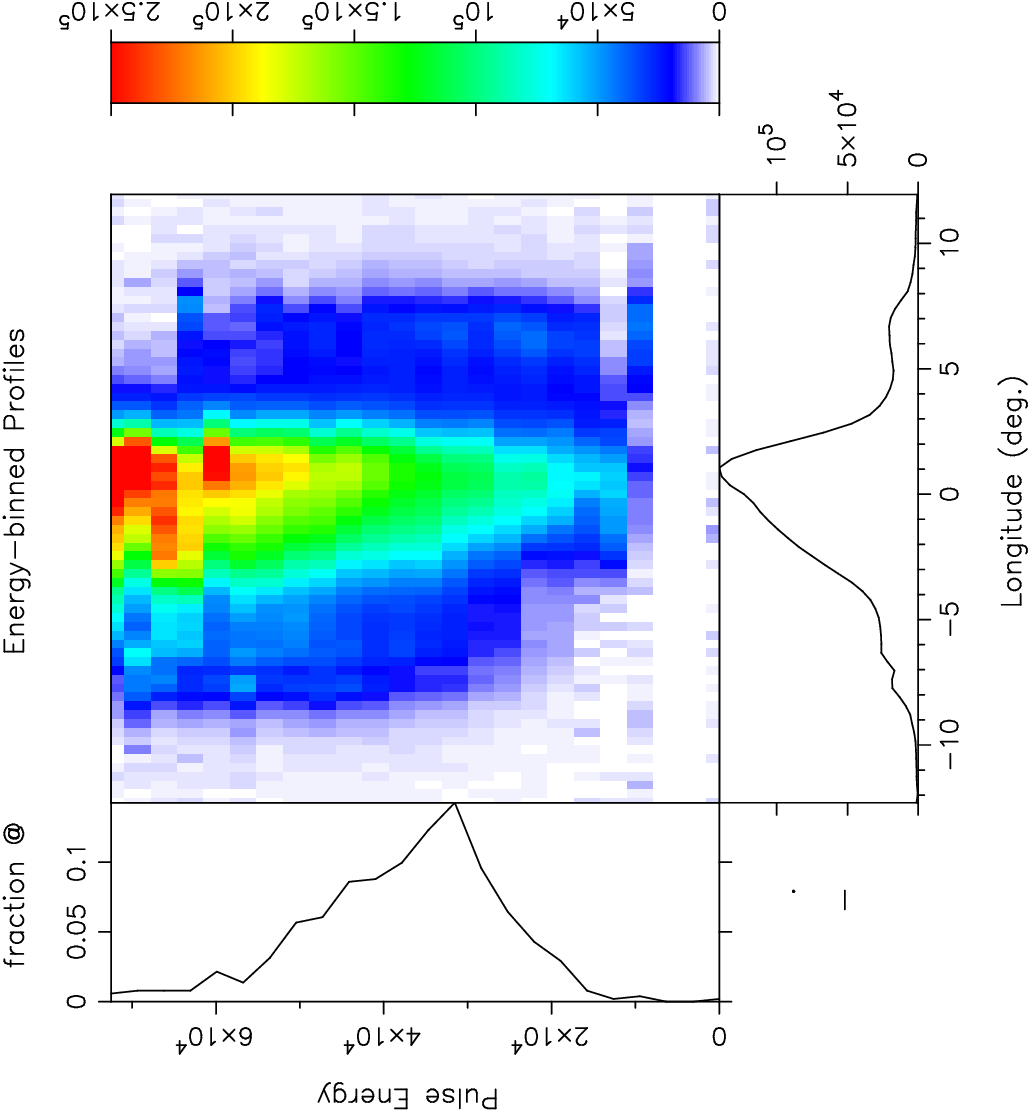}}&
\mbox{\includegraphics[height=70mm,width=60mm,angle=-90.]{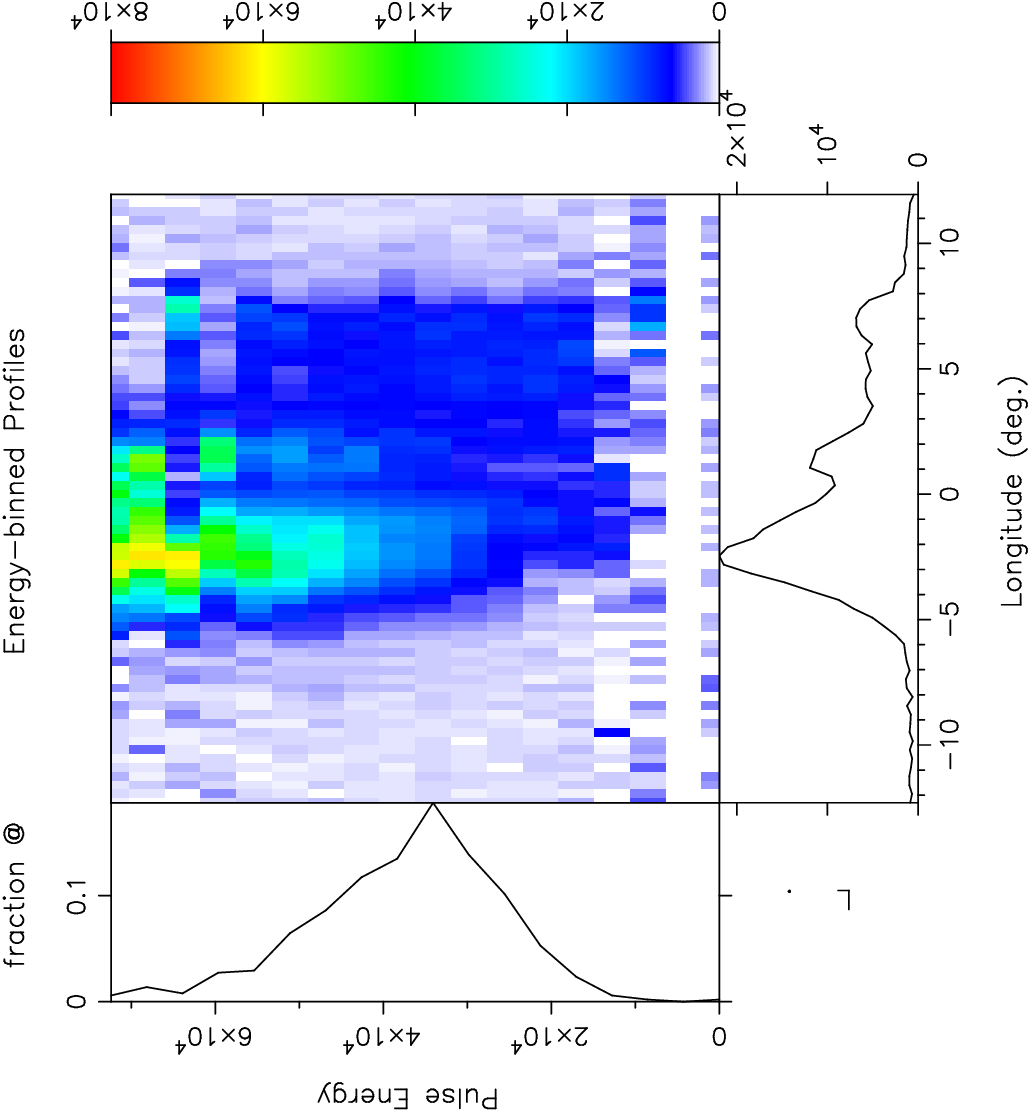}}\\
\end{tabular}
\caption{Intensity-fractionated displays for the LSP3 observation in Fig.~\ref{fig1} upper 
right:  total power $I$ (left) and total linear $L$ (right).  Note the pronounced shift to 
earlier longitudes of the higher intensity emission.  This intensity-dependent 
aberration/retardation is responsible for the strange non-RVM ``hook'' in the PPA 
distribution between about --7 and --1\degr\ longitude.  The effect is seen in $L$ as well 
as prominently in $I$ and is responsible for the unresolved double shape of the core 
component.}
\label{fig2}
\end{figure*}

\section{Arecibo Observations}
\label{obs}

The single pulse polarized pulse sequences for PSR B1946+35 were obtained with 
the 305-m Arecibo radio telescope in Puerto Rico together with its Gregorian reflector 
system and both the L-band-wide and C-band antenna-receiver systems.  
The observations were part of larger proposals which
aimed to study the emission mechanism using single-pulse polarization 
observations of a large number of pulsars
across multiple frequencies. The MJD and the Arecibo project ID for the
observations used here are given in Table~\ref{tab1}.
The L-band 
pulse sequence observed on 17 July 2003/MJD 52837 (hereafter referred to as LPS1) was obtained 
using four WAPP (Wideband Arecibo Pulsar Processor) spectrometers, and the 12 March 2011/MJD 55632 and 
7 September 2016/MJD 57638 (hereafter LPS2 and 
LPS3) observations used four Mock spectrometers with nominal 100- and 86-MHz 
bandwidths, respectively.  The C-band pulse sequence (hereafter referred as CPS1) 
was observed on 19 September 2015/MJD 57284 using the seven Mock spectrometers with adjacent 
170-MHz bands each.  For each observation the raw Stokes parameters obtained for 
each band were corrected for dispersion and interstellar Faraday rotation. At 
L-band 
the three lower bands were found relatively free from interference and were thus
added together to give an effective roughly 300-MHz bandwidth, whereas at C-band  
3 bands were omitted owing to interference, giving an effective usable bandwidth of 
$\sim$ 600 MHz.  Other relevant observing parameters are given in Table~\ref{tab1}.
Fig.~\ref{fig1} shows the single pulse sequence and polarization for the LPS3 and 
CPS1 observations.  The L-band pulse sequence clearly shows periodic modulation 
for the central core and trailing conal component, while the modulation in the leading 
conal component is most clearly seen in the C-band observation.

\section{Classification and Quantitative Geometry}
\label{sec3}
Pulsar B1946+35 was classified as having a core-single ({\bf S$_t$}) profile in ET VI, 
and its emission geometry was there modeled according to the RVM and core/double-cone 
models.  Here, we review in detail the basis of this classification using all the available 
lines of evidence.

\subsection{Profile and Spectral Evolution}
Pulsar B1946+35 has a dispersion measure $DM$ of 129.07 ${\rm pc\,cm^{-3}}$, and its profile 
shows substantial scattering at meter wavelengths.  Hence, only observations at higher 
frequencies are useful for studying the intrinsic emission properties of this pulsar.  The 
spectral evolution of the average total power profile of PSR B1946+35 was published by 
\citet{1988MNRAS.234..477L} for 0.6, 1.6 and 2.6 GHz, and a 4.8-GHz profile appears in 
\citet{1997A&AS..126..121V}.  Our observations at L-band and C-band (see Fig.~\ref{fig1}) 
have a significantly higher signal-to-noise ratio and are consistent with the earlier results.
The primary characteristic of core-single ({\bf S$_t$}) profile evolution is clearly seen in 
these observations.  The pulsar's profile has but a single bright component at 0.6 GHz; 
however, with increasing frequency two roughly symmetrical conal outriding components 
become ever more prominent until they are dominant at 4.8 GHz.  Using our own 1.4-GHz 
observations, we estimate the half-power width of the central core component to be about 
$W_c$ = 5.0\degr$\pm$0.05\degr.  The \citet{1998MNRAS.301..235G} profiles provide a width measurement 
at 0.92 GHz of 5.4\degr$\pm$0.1\degr, and interpolating between 0.92 and 1.4 GHz, we 
obtain a 1-GHz core-component width of 5.2\degr$\pm$0.1\degr.  Thus the geometrical 
interpretation of relating the core width to the polar cap size gives $\alpha \sim $32\degr.  
This is consistent with the $\alpha$ value estimated in papers ET IV and VI for this pulsar 
using core-width measurements from \citet{1989ApJ...346..869R}. 


\begin{figure}
\begin{center}
\begin{tabular}{c}
\mbox{\includegraphics[height=80mm,width=90mm,angle=-90.]{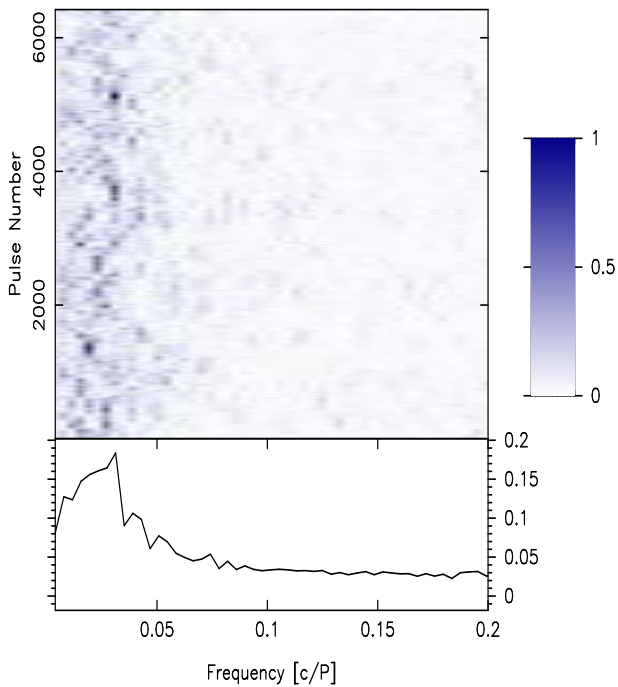}} \\
\end{tabular}
\caption{Display showing the time variations of the LRFS within the LPS3 pulse 
sequence.  The top panel is a colour-coded representation with the longitude-averaged 
LRFS along the x-axis and increasing starting point along the y-axis.  In order to closely 
show the observed spectral variations, the fluctuation scale is plotted only up to 0.2 c$P^{-1}$ 
(rather than the usual 0.5 c$P^{-1}$).  The bottom panel shows the aggregate LRFSs along 
the y-axis (see \S\ref{sec4} for details)}
\label{fig3}
\end{center}
\end{figure}

\subsection{Polarimetry}
The PPA histograms seen in the right-hand displays of Fig.~\ref{fig1} for the 1.4-GHz 
LPS3 and 4.6-GHz CPS1 observations appear at a first glance to be quite complex 
and inscrutable in RVM terms.  They are complicated by orthogonal and non-orthogonal 
PPA modal jumps and swings.  However for the LPS3 data, if the region between --7 and 
--1\degr\ longitude is ignored, a reasonable RVM traverse can be obtained that includes
an orthogonal jump at +6\degr\ longitude.  Mindful of the large correlation in fitting for 
$\alpha$ and $\beta$, we fitted instead for the steepest gradient point slope $R_{\rm pa}$ 
[$= \sin\alpha/\sin\beta$], obtaining a PPA rate of 16$\pm$3\degr/\degr\ and a PPA at 
the inflection point $PA_0$ of --66\degr$\pm$5\degr, both of which have well determined 
errors. The longitude origin of the lower plot in Fig.~\ref{fig2} corresponds to this inflection
point and is also consistent with the point where the circular polarization changes sign.  
Using the $\alpha$ value of  $\sim$32\degr\ from the core-width measurement together 
with the above $R_{\rm pa}$ value, we computed $\beta$ as 1.9\degr$\pm$0.3\degr.   For 
the CPS1 observation in Fig.~\ref{fig1} the single pulses are weaker; however the PPA 
histogram is well sampled around --10\degr\ to --8\degr\ and 4\degr\ to 7\degr\ longitude 
showing that the values lie in approximately the same locations as in the LPS3 profile.  
Since the PPAs in these displays are derotated and thus reflect absolute orientations 
on the sky, the measurements indicate that the PPAs are frequency independent.  The 
RVM measured for the upper LPS3 1.4-GHz observation, which is then imposed on the 
lower CPS1 4.6-GHz profile, then seems to be in good agreement with its average PPA 
traverse.   

The steep PPA swing between --7 and --1\degr\ longitude is a curious feature that 
initially perplexed us.  However, on closer inspection we noticed that the core feature 
is comprised of two blended, roughly Gaussian-like structures which together result 
in the asymmetric component.  Analysis using intensity fractionation then revealed 
the intensity-dependent aberration/retardation that is responsible for the PPA ``hook'' 
as seen in Fig.~~\ref{fig2}.  Similar PPA anomalies have now been identified and 
studied in the leading regions of bright core components of pulsars PSR B0329+54, 
B1237+25 and B1933+16 \citep{2007MNRAS.379..932M,2013MNRAS.435.1984S,2016MNRAS.460.3063M}.
There is some indication that the swing is produced 
by unequal mixing of orthogonal polarization-mode power occurring within that longitude 
range, however we were not able to make this case definitively for PSR B1946+35.  

We now turn to an analysis of the pulsar's basic emission geometry.  Using the above 
$\alpha$ and $\beta$ values together with the measured widths at the outer half-power 
points of the conal component pair, we can compute the conal beam radius $\rho$. 
Following the spherical trigonometry methods of paper ET VI (eq. 4), this results in a 
value of approximately 4.6\degr\ around 1 GHz (the measured width and $\rho$ for 
all the observations are given in Table~\ref{tab1}). This in turn is compatible with an 
inner cone geometry for a pulsar having a rotation period of 0.717 s 
\citep[See also e.g.][]{1984A&A...132..312G}.
Note in addition that $\rho$ is practically constant 
between L-band and C-band, which is also an identifying inner cone property (ET VI \& VII).

Thus, in summary, we have reestablished that pulsar B1946+35's profile dimensions 
and polarization are consistent with a core-single ({\bf S$_t$}) classification wherein 
the outriding conal components correspond to an inner cone geometry and the central 
component is well identified as a core emission feature. Assuming the outer half-power 
points to lie along last open field lines, one can define a parameter $s$ as the locus of 
the dipolar field line compared to polar fluxtube boundary such that $s=1$ corresponds 
to the edge of the open field region and $s=0$ is the beam center.  Using the fact that 
the annular width of the cone is about 20\% of the beam radius, the range of field lines
over which the inner cone is illuminated in the beam is from $s \sim 1$ to $0.8$. 
Correspondingly the core emission illuminates field lines inwards of $s \sim 0.45$.
This in turn ensures that the observer's sightline through the beams is such that the 
core emission is relatively unaffected by any overlapping conal emission power.

\begin{figure*}
\begin{center}
\begin{tabular}{cc}
\mbox{\includegraphics[width=70mm,angle=-90.]{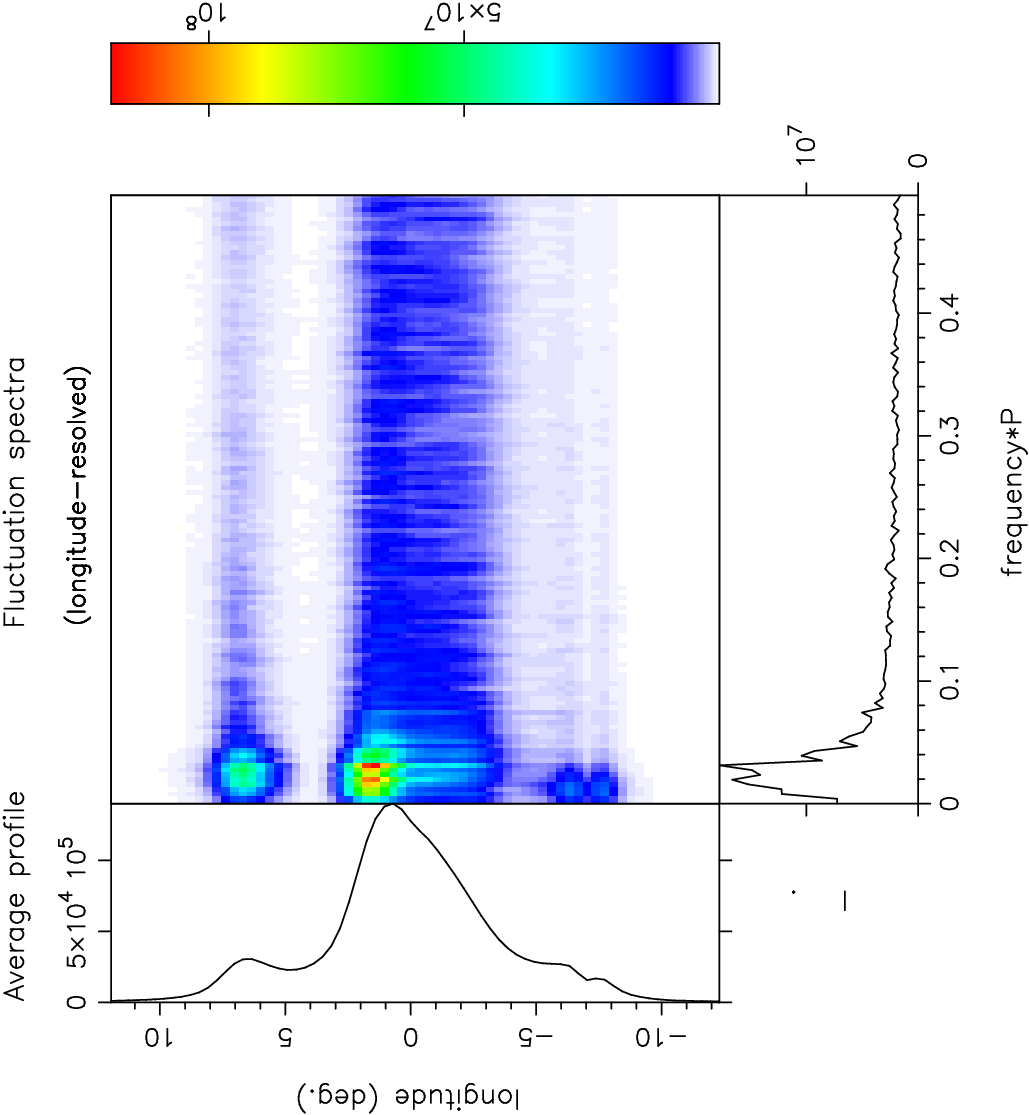}}&
\mbox{\includegraphics[width=70mm,angle=-90.]{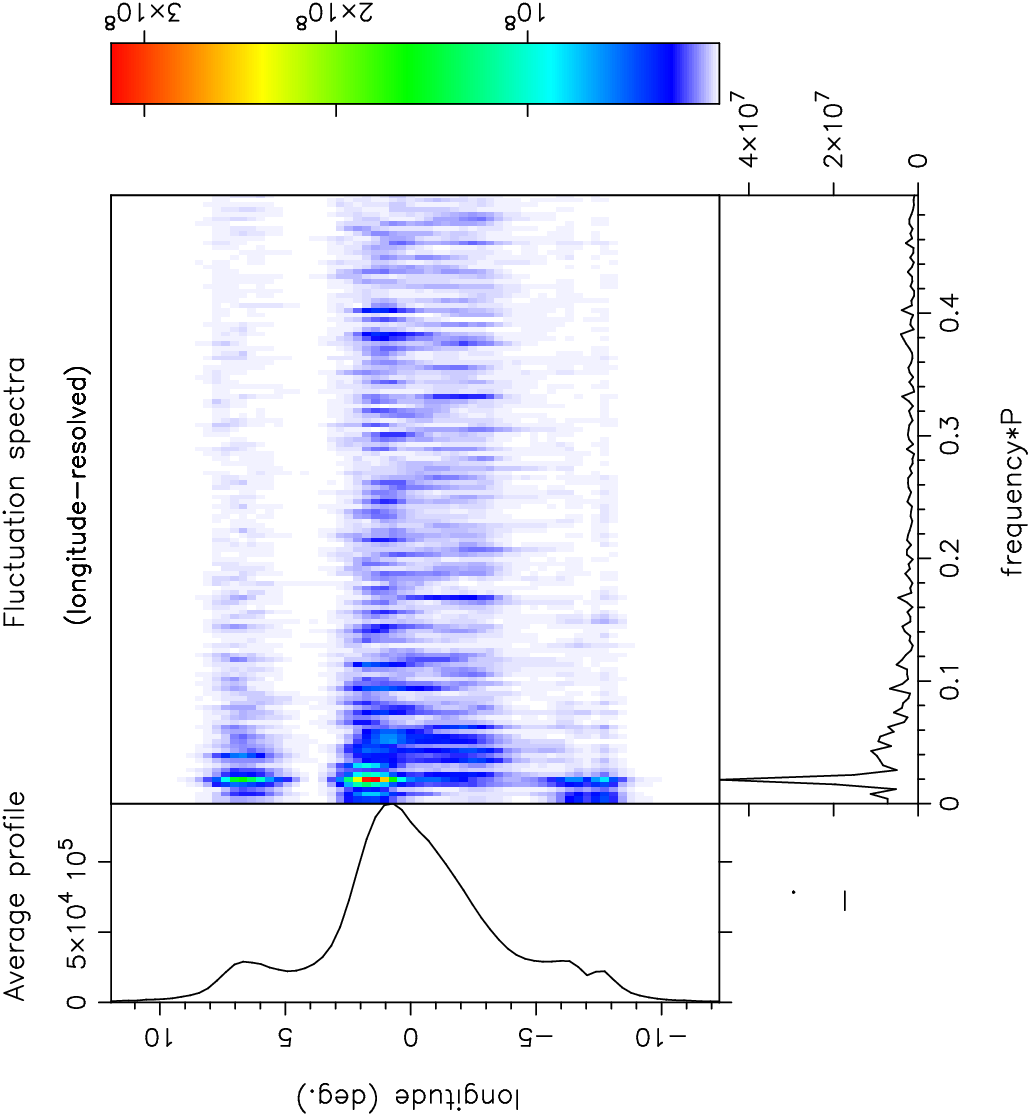}}\\
\mbox{\includegraphics[width=60mm,height=70mm,angle=-90.]{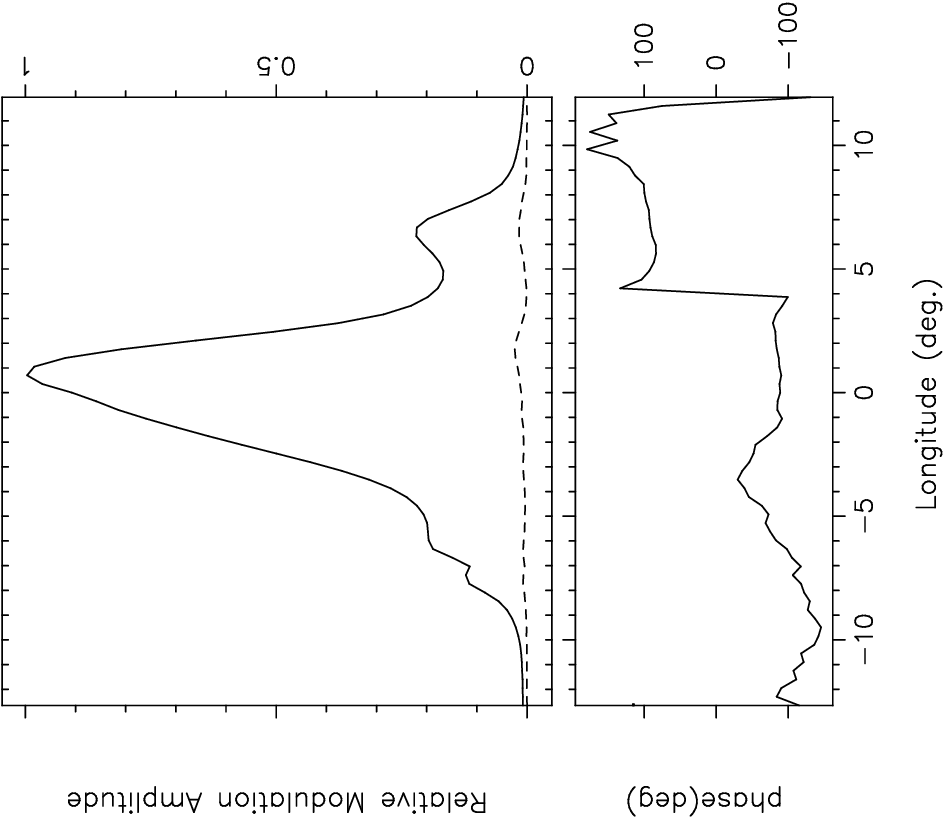}}&
\mbox{\includegraphics[width=60mm,height=70mm,angle=-90.]{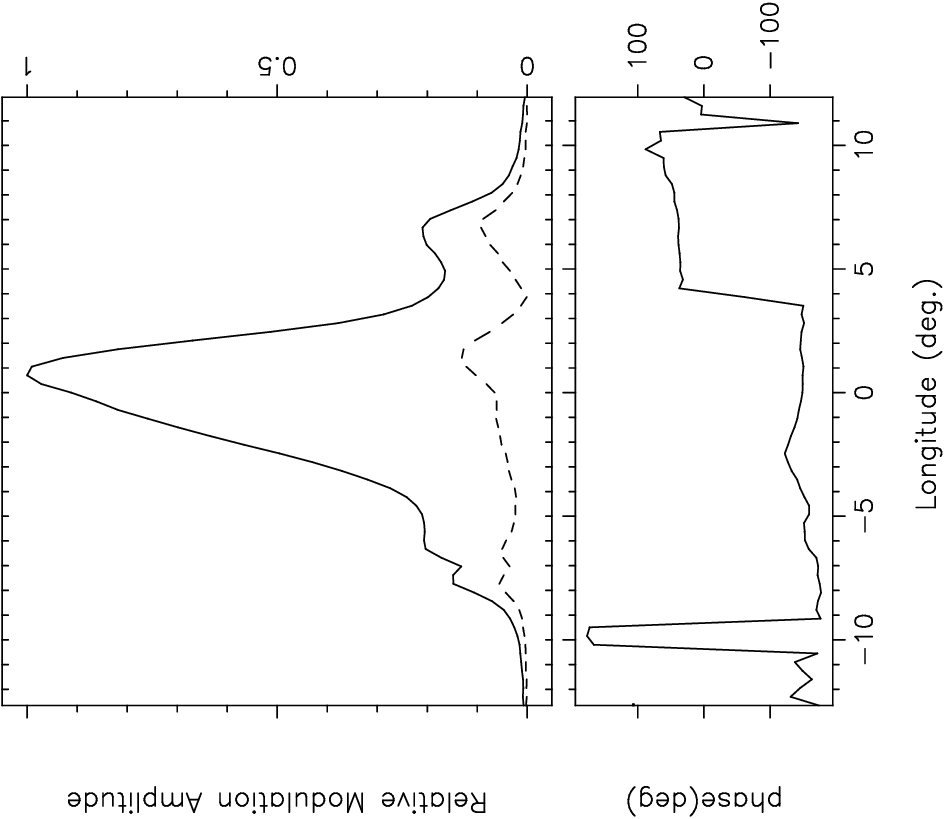}}\\
\mbox{\includegraphics[width=70mm,angle=-90.]{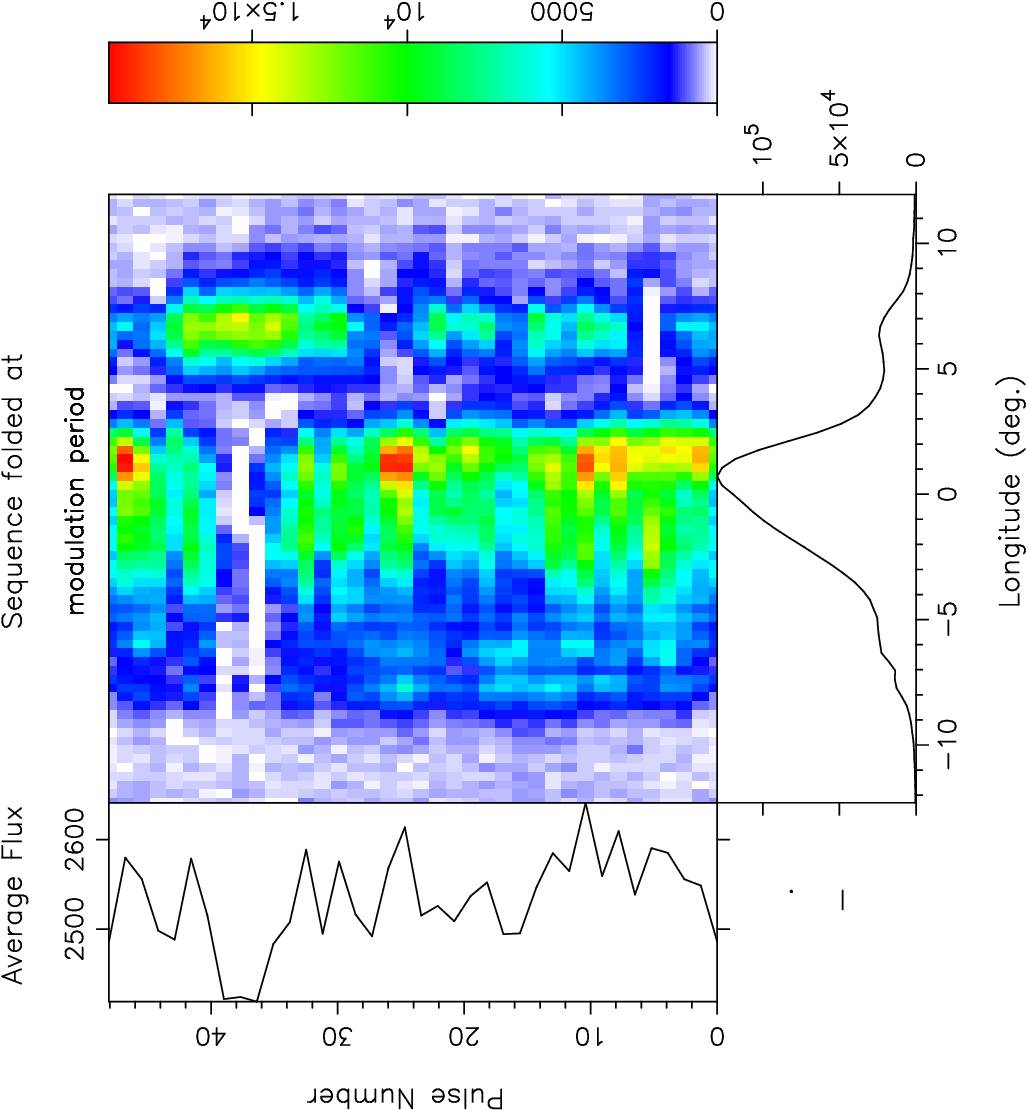}}&
\mbox{\includegraphics[width=70mm,angle=-90.]{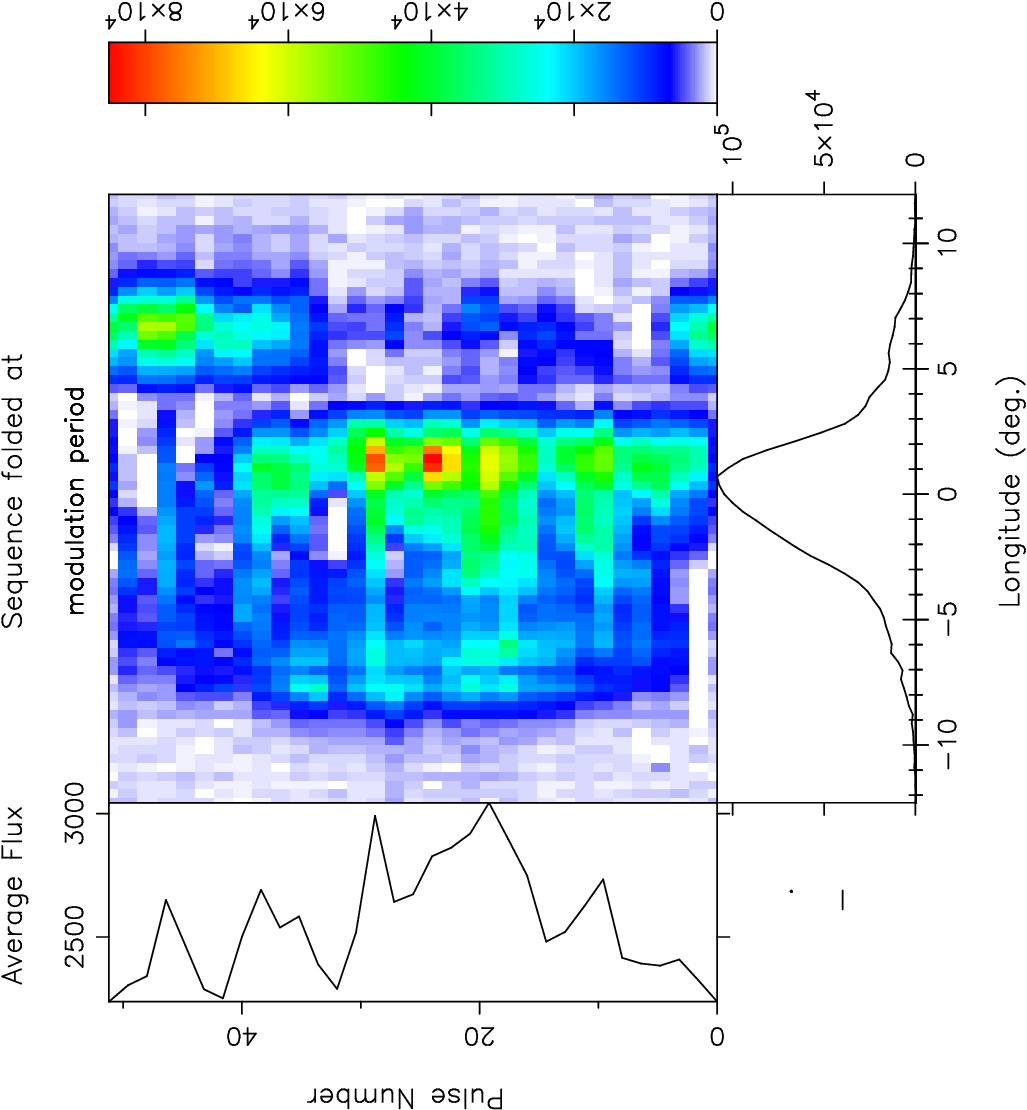}}\\
\end{tabular}
\caption{Total intensity, longitude-resolved fluctuation spectra using overlapping 
FFT lengths of 256 points (upper displays), modulation-phase plots (middle displays), 
and profiles folded at the modulation period (bottom) are given for the whole LPS3 
observation (left) and for the 1250--1761 pulse interval (right).  In the upper displays 
the main-panel fluctuation intensity is colour-coded according to the scale at the right, 
the average profile is given on the left and the aggregate intensity in the lower panel.  
The central displays show the average profiles, the fraction of modulated power 
(dashed curve) and the phase function (lower panels).  Displays at the bottom show 
the character of profile changes over the modulation cycle (main panel, colour-code 
per the scale at the right); the total power fluctuation is shown at the left and the constant 
``base'' profile at the bottom (see \S\ref{sec4} for details).}
\label{fig4}
\end{center}
\end{figure*}

\section{Fluctuation Spectral Analysis}
\label{sec4}
A clear suggestion of periodic fluctuations in PSR B1946+35 can be seen in the 
single pulse sequences of Fig.~\ref{fig1} both at L-band and C-band.  This was also 
implicit in the fluctuation spectra analyses of WES06 and WES07. Such behavior 
is unusual for a core-single pulsar.  In order to investigate the character of 
these fluctuations in more detail we proceeded to compute longitude-resolved 
fluctuation spectra (hereafter LRFS) for this pulsar in a number of different ways. 
WES06 and WES07 found two different $f_3$ values using short intervals of P- 
and L-band observations. We wondered if this difference could result from a 
time-varying fluctuation property of the pulsar, and this prompted us to first look 
for temporal variations in the long LPS3 sequence using techniques similar to 
those described by \citet{2016ApJ...833...29B}.  
In this method LRFSs of 256 pulses were computed as a function of time after 
shifting the starting pulse successively by ten periods. For every time realization, 
the LRFSs were averaged across pulse longitude such that a single fluctuation 
spectrum was obtained, and finally after shifting each by ten periods these LRFSs 
were represented on a two dimensional map.  This map was normalized 
to unity and is shown as a colour-coded distribution in the top panel of Fig.~\ref{fig3}, 
where the ordinate corresponds to the starting pulse for computing LRFSs and 
the abscissa to the LRFSs frequency in c$P^{-1}$.  The  display 
in the bottom plot of Fig.~\ref{fig3} shows the average LRFS after collapsing all 
the LRFSs along the vertical axis.  

Remarkably, an interesting pattern was observed in the temporal variations of 
the LRFS, where for about thousand pulses the LRFS shows a broad low 
frequency ``red'' power excess, but it then suddenly exhibits a very different 
behavior for a few hundred pulses characterized by a highly periodic modulation.  
We looked for this effect in the shorter LPS1, LPS2 and CPS1 observations, and 
found intervals having similar broad low frequency excesses and also sections
with highly periodic modulations.  To further quantify this effect we resorted to 
analyzing sections of the observations by performing overlapping 256-point 
FFTs (see \citealt{2001MNRAS.322..438D} for the method). We found that the LRFS 
for each entire pulse sequence revealed the broad low frequency excess, and 
we extracted a characteristic fluctuation frequency by fitting the parametric 
B\'ezier curve to the average LRFS and locating the peak frequency of this fitted 
curve which we will refer to as $f_3^{offt}$.    

For highly modulated periodic intervals of the pulse sequences we fitted a cubic 
spline to the sharp spectral feature to find the corresponding peak $f_3^{offt}$ 
values.  The sharpness of the spectral feature was estimated by computing a 
``quality factor'' $Q = f_3^{offt}/\Delta f$, where $\Delta f$ corresponds to the 
half-power width of the feature.  The range of pulses over which the FFTs were 
computed, the number of points for the overlapping FFTs, values of $f_3^{offt}$ in c$P^{-1}$, and Q factors for all the sequences are quoted in Table~\ref{tab1}. The 
errors in $f_3^{offt}$ were estimated using eq.(4) of \citet{2016ApJ...833...29B}. 
Note that these highly modulated features appear at three distinct
frequencies, although they are not harmonically related to each other. 

The LRFS analysis described above reveals $f_3^{offt}$ values in sections of the 
observations which are similar to the two different values found by WES06 and 
WES07 at two different observing frequencies.  For example two pulse ranges 
from the LPS3 observation, one in the interval 3550--4062 had 
$f_3^{offt}$=0.0312$\pm$0.003 c$P^{-1}$, which is comparable to the L-Band 
WES06 value of 0.031 (their $P_3$=$33\pm2$$P$), while another interval 
1250--1761 had $f_3^{offt}$=0.0193$\pm$0.003 c$P^{-1}$, which is similar to the 
WES07 P-band value of 0.018 (their $P_3$=$55\pm9$$P$).  Thus we conclude 
that the variation in the modulation frequency reported earlier is not  
frequency dependent, but rather occurs at different times when the pulsar 
is observed at any frequency. 

The intervals of periodic modulation in the observations are seen not only in the 
two conal components, but also in regions on both sides of the core component.  
In Fig.~\ref{fig4} we show the results of detailed LRFS analyses of the full LPS3 
pulse sequence (left column) and the pulse range 1250--1761 (right column).
The top panels show the LRFS, the middle displays plot the amplitude and 
phase of the spectral feature across the longitude range of the pulse profile, and 
the bottom panels show the pulse sequence folded at the $f_3^{offt}$ frequency 
as given in Table~\ref{tab1}.  Notice that in the LRFS for the entire pulse 
sequence the average spectral feature (lower panel) is broad and diffuse in nature, 
but is clearly seen to fluctuate in both the core and conal profile regions.  However, 
within the specific interval where the pulse sequence displays periodic modulation, 
the narrow spectral feature $f_3^{offt}$ is clearly seen across the profile.  

The peak signal-to-noise ratio (S/N) of the feature is about 100 (the noise being 
calculated in higher frequency regions between 0.3 to 0.35 c$P^{-1}$ where the spectum 
is featureless and dominated by white noise), and the power 
up to the half-power point is confined to a single FFT frequency component giving 
a factor $Q=f_3^{offt}/\Delta f \sim 3.2$ (see Table~\ref{tab1}).  The middle displays 
show the average amplitude and phase of the spectral feature across the longitude 
range of the pulse profile.  The amplitude under the leading component shows the 
least fluctuating power, whereas the trailing part of the core component is more 
strongly modulated than the leading part and is similar in magnitude to that of the 
trailing component.  The modulation phase is remarkably constant across each 
component implying that the fluctuations correspond to pure amplitude modulation.
The leading conal component and the core region are modulated very closely in 
phase, whereas the trailing component is modulated almost precisely in counter 
phase.  These various aspects of the periodic modulation can also be seen in the 
bottom displays of Fig.~\ref{fig4}, where the pulse sequence has been folded at 
the $f_3^{offt}$ frequency.  Here we see that for a 24-25$P$ half-portion of the 
modulation cycle, the pulsar's leading conal and core components are active, and 
then for the remaining half of the cycle, these weaken and the bright trailing conal 
component appears and stays active.  The corresponding harmonic-resolved 
fluctuation spectrum (HRFS, see \citealt{2001MNRAS.322..438D,2016ApJ...833...29B}) which lies
in 0--1 c$P^{-1}$ range 
provides an alternative way of determining the phase behaviour of the fluctuating
feature. The feature in the HRFS (not shown) lies symmetrically about 
the center point 0.5 c$P^{-1}$ which indicates that the 
responses represent largely amplitude modulation as expected.

\section{Comparison with Other Pulsar Modulation Phenomena}
\label{sec5}
Over time scales of few thousand pulses radio pulsars are well known to exhibit 
three major types of pulse-sequence effects: subpulse drifting, pulse nulling, and 
mode-changing (\eg see ET III).  Amongst these, high Q periodic modulation is 
primarily seen in the subpulse-drifting phenomenon.  The periodicities found are 
typically longer than the pulsar rotation period $P$ and can be as long as a few 
tens of seconds.  The fluctuation-spectral analyses described above for B1946+35 
suggest periodic modulation of about 35 s, and we now contrast its properties 
with the subpulse-drift phenomenon. 

High Q drift features are seen primarily in conal single {\bf S$_d$} pulsars, where 
the fluctuation is associated with distinct phase modulation across the pulse profile. 
In some situations high Q periodic stationary or amplitude modulation is observed in conal 
component pairs [\eg PSR B1857--26 \citep{2008MNRAS.385..606M}; B1237+25 
\citep{2013MNRAS.435.1984S} 
and B2045--16 \citep{2016ApJ...833...29B}] which correspond to conal double 
profiles with central sightline traverses.  
In all these cases however, the central 
core emission never seems to show any high Q periodic modulation.  This 
geometrical property of the drifting-subpulse phenomenon has given rise to 
the carousel model, wherein the drifts of the conal emission are understood as 
due to a persistent system of localized beamlets which rotate on a roughly circular 
path around the dipolar magnetic axis.  The core emission is associated with a 
beam which is anchored near the dipole axis and hence is phase stationary.  We 
have questioned whether the regular modulation in B1946+35 might be due to some 
kind of carousel-beam system, and such a mechanism for the conal fluctuations 
cannot be entirely ruled out.  However, the joint modulation of B1946+35's well 
identified core emission together with that of the cones is unusual and paradoxical 
here.  If we insist that the observed modulation results from moving beamlets that 
cross our sightline, then both the conal and core beamlets should be moving 
perpendicular to the sightline circle---and thus in contradiction with the carousel 
model.

The physical motivation for the carousel model was outlined by RS75, 
where they suggested that an inner vacuum gap (IVG) exists 
near the pulsar polar cap which discharges as a set of the localized sparks.
Through a complex process of electron-positron pair creation, a spark-generated
relativistic plasma flow is established which gives rise to the radio emission at 
altitudes of a few hundred kilometers above the neutron star surface.  Hence 
each spark is associated with a plasma column and appears as a beamlet in 
the emission region.  Due to {\bf E}$\times${\bf B} drift in the IVG, the sparks 
experience a slow (magnetic azimuthal) drift which lags the pulsar corotation.  
As a result the radio emitting beamlets also lag the corotation, and this gives 
rise to the subpulse drifting phenomenon.  In an LRFS analysis, the modulation 
frequency $f_3^{offt}$ can be interpreted in terms of the spark repetition time 
$P_3$.  RS75 explored this problem for an antipulsar where the rotation and 
magnetic axes are aligned and hence the beamlet motion is both around the 
rotation and magnetic axes---and this is the genesis of the carousel model.  
Application of this model to actual pulsars---where their rotation and magnetic
axes are necessarily non-aligned---is far from trivial.  While there has been 
observational support for the carousel model wherein the sparks rotate around 
the magnetic axis, \citet{2016ApJ...833...29B} have recently pointed out that the model 
is inconsistent physically with the fundamental circumstance that sparks must 
lag behind a pulsar's corotation.  

The lack of any significant periodic modulation in pulsar core emission has 
been the most stringent observational constrain on the carousel model, since 
the core has been interpreted as a stationary (non-drifting) beam located along 
the dipolar magnetic axis.  However, the modulation we observe in the core of 
B1946+35 breaks that constraint.  One key question then naturally arises:  can 
some refinement of the RS75 spark-drifting model explain the host of drifting 
phenomena observed, including both the phase and amplitude modulation of 
periodic fluctuations.  Recently, in an effort to resolve this problem, 
\citet{2016ApJ...833...29B} and \citet{2013arXiv1304.4203S} 
suggested that if the magnetic field 
in the IVG is highly non-dipolar, then the motion of sparks lagging the corotation 
in the IVG can produce both phase- and amplitude-stationary modulation in the 
dipolar radio-emission region.  However, it is too early to conclude whether 
this idea can explain the broad range of observed subpulse-drift properties.

There is also another aspect of drifting that needs attention while interpreting 
the PSR B1946+35 modulation.  Recently \citet{2016ApJ...833...29B} carefully applied 
the RS75 conception that sparks lag behind pulsar corotation, and they found 
that pulsars with phase- and amplitude-modulated drifting could be seen as two 
separate populations in $\dot{E}$ space:  One group of pulsars with with both 
phase and amplitude drift modulation are seen to lie below $\dot{E}\sim$ 2$\times$10$^{32}$ 
${\rm ergs\,s^{-1}}$, and the drift periodicity $P_3$ was found to be anti-correlated with 
$\dot{E}$ roughly as $P_3 \sim$ ($\dot{E}/2$$\times$10$^{32})^{-0.6} P$; whereas 
the other group for which $\dot{E} >$ 2$\times$10$^{32}$ ${\rm ergs\,s^{-1}}$ all showed only 
amplitude modulation with a median value of $P_3 \sim$ 30$\pm$15 $P$.
\citet{2016ApJ...833...29B} interpreted this fascinating result using a partially screened 
vacuum gap or PSG model \citep{2003A&A...407..315G}---which is the modified RS75 model 
wherein the IVG is partially screened due to presence of ions that are extracted 
from the neutron-star surface.  They demonstrated that under the PSG model, the 
$P_3$ {\it vs.} $\dot{E}$ anticorrelation can be explained as an increase in the spark 
drift speed with increasing $\dot{E}$ of the pulsar. This also means that above a 
certain $\dot{E}$, when $P_3$ becomes less than the $P$, subpulse-drift motion 
is aliased and cannot any longer be observed.  This then raises the question as 
to what is the physical origin of the phase-stationary amplitude modulation seen 
in the group two pulsars.  PSR B1946+35 has an $\dot{E} \sim$ 7.6$\times$10$^{32}$ 
${\rm ergs\,s^{-1}}$, and lies in group two.  Its phase-stationary periodic modulation could then 
be an entirely new phenomenon.

Similarly, the phenomenon known as ``mode changing''---wherein a pulsar's average 
profile assumes several different discrete forms and the individual pulse properties 
so change to produce them---is well documented in a variety of different pulsars.  
In several of the early exemplars of mode-changing (\eg B0329+54 and B1237+25) 
the effect took the form of a change in the symmetry of the conal component emission 
about the central core component, and this description surely could be taken to apply 
to B1946+35 as well, although in this case the changes occur on a significantly
shorter time scale of about 18 s.  The other problem is that mode-changing is not 
known to be periodic, and surely not periodic with a reasonably high $Q$ as we 
see in this pulsar.  

The high Q drift feature in pulsars is known to show gradual changes
across a particular mode \citep[e.g. B0943+10, see][]{2006A&A...453..679R} or when 
the pulsar recovers from its nulling phase (e.g. B0809+74 \citealt{2002A&A...387..169V}).  
In certain drifters like B0031--07 \citep{1997ApJ...477..431V,
2005A&A...440..683S,2017arXiv170106755M}, B1944+17 \citep{2010MNRAS.408...40K}, 
B1918+19 \citep{2013MNRAS.433..445R} and B2319+60 \citep{1981A&A...101..356W} 
the high Q modulation feature is seen to have multiple (often three) distinct
high Q modulation frequencies often interspersed with long duration nulling 
episodes. In this context it is interesting to note that PSR B1946+35
switches into a high Q mode after spending long intervals
in a mode marked by low frequency excess. For the long L-band observation, 
we found that the pulsar can settle into three distinct high Q periodicities as 
seen in Fig.~\ref{fig3} and Table~\ref{tab1}.
 
Indeed, overall it is conal emission that exhibits well established periodicities, 
not cores.  Core radiation more typically presents as ``white'' noise in flat or 
featureless fluctuation spectra.  A few examples of long broad periodicities 
in core components are known (\eg \citealt{1986ApJ...301..901R},WES06, WES07), 
though very few have been studied in detail and documented.  In pulsar  
B1237+25, the core-associated periodicity seems to be associated with its three 
modes \citep{2005MNRAS.362.1121S}, where in the ``quiet normal'' mode core 
``flares'' are observed at rough intervals of some 60 periods.  Core eruptions also 
seem responsible for the ``red noise'' (noted in the first paper above) in the central 
regions of several classical conal double pulses, though no clear periodicities 
have so far been established \citep{2012MNRAS.424.2477Y}.  
It begins to seem that core fluctuations might be a more widespread phenomenon
and needs further investigation.

Recently, several new quasiperiodic bursting phenomena, akin to mode changes, 
have been identified in radio pulsar emission---\eg  PSR J1752+2359 
\citep{2004ApJ...600..905L}; J1938+2213 \citep{2013MNRAS.434..347L}; B0611+22 \citep{2014MNRAS.439.3951S}.  For PSR B0611+22  \citet{2016MNRAS.462.2518R} 
explored whether there might be simultaneous X-ray emission changes during its radio 
mode changes. However, they did not detect X-ray emission from this pulsar, and 
hence the mode changes could not be associated with magnetospheric ``states'' as 
have been observed for the radio and X-ray mode changing pulsar B0943+10 
\citep{2013Sci...339..436H,2016ApJ...831...21M}. Another quasiperiodic phenomenon 
has been studied in pulsars B0919+06 and B1859+07 with bistable behaviour 
\citep{2006MNRAS.370..673R,2016MNRAS.456.3413H,2016MNRAS.461.3740W}, 
wherein occasionally the bright emission window shifts earlier for a few or several 
score periods and then returns to its usual phase.  This resembles in several ways the emission 
pattern seen in PSR B1946+35, where for about 25 $P$ the pulsar's emission window 
corresponding to the leading conal and core component appears shifted, followed 
by an active phase of similar duration for the trailing conal component.  However, 
the timescales of these bursting or bistable phenomena 
\citep[see also][]{2010Sci...329..408L} are larger or much larger than the B1946+35 
modulation period.  Interestingly, \citet{2016MNRAS.461.3740W} explore whether the 
bistable quasi-periodicity could arise due to the emission being affected in a periodic 
manner by a `satellite' object rotating in a binary system.  Given the observed 
modulation periodicities of about 300 s and 100 s in B0919+06 and B1859+07, they 
found that the semi-major axis of the companion should be at a distance comparable 
to the light cylinder and have densities larger than $10^5$ ${\rm gm\,cm^{-3}}$ to avoid tidal 
disruptions. We however do not advocate such a mechanism for PSR B1946+35, 
with a modulation period of only some 36 s, thus making the orbit fall well within 
the light cylinder.

\section{Conclusion}
\label{sec6}
As argued above the highly periodic phase-stationary fluctuations in intervals of 
pulsar B1946+35's modulation cannot be reconciled with the drifting-subpulse 
phenomenon in any conceivable geometrical context. The periodic fluctuations 
in this pulsar appear to require some sort of emission-pattern changing on timescales 
of about 18 s.  How this new effect may or may not be related to the well known 
mode-changing phenomenon remains unclear.  Mode changes have generally 
been documented through use of modal profiles computed over much longer time 
intervals, and periodicities have not heretofore been documented as an aspect of 
this phenomenon.  Longer duration observations at multiple frequency bands will 
throw more light on the nature of this new phenomenon. 

\section*{Acknowledgments} We thank the anonymous referee for 
comments and suggestions which significantly improved the paper. 
We thank Rahul Basu for several intersting discussions. 
Much of the work was made possible by support from a Vermont Space Grant
and NSF grant 09-68296.  Arecibo 
Observatory is operated by SRI International under a cooperative agreement with the 
US National Science Foundation, and in alliance with SRI, the Ana G. M\'endez-Universidad 
Metropolitana, and the Universities Space Research Association.  This work made use 
of the NASA ADS astronomical data system.

\bibliographystyle{mnras}
\bibliography{B1946+35_final}
\end{document}